# Are the discretised lognormal and hooked power law distributions plausible for citation data?

Mike Thelwall, Statistical Cybermetrics Research Group, University of Wolverhampton, UK.

There is no agreement over which statistical distribution is most appropriate for modelling citation count data. This is important because if one distribution is accepted then the relative merits of different citation-based indicators, such as percentiles, arithmetic means and geometric means, can be more fully assessed. In response, this article investigates the plausibility of the discretised lognormal and hooked power law distributions for modelling the full range of citation counts, with an offset of 1. The citation counts from 23 Scopus subcategories were fitted to hooked power law and discretised lognormal distributions but both distributions failed a Kolmogorov-Smirnov goodness of fit test in over three quarters of cases. The discretised lognormal distribution also seems to have the wrong shape for citation distributions, with too few zeros and not enough medium values for all subjects. The cause of poor fits could be the impurity of the subject subcategories or the presence of interdisciplinary research. Although it is possible to test for subject subcategory purity indirectly through a goodness of fit test in theory with large enough sample sizes, it is probably not possible in practice. Hence it seems difficult to get conclusive evidence about the theoretically most appropriate statistical distribution.

## 1. Introduction

Citation-based indicators are sometimes used to support formal evaluations of groups of academics and for self-evaluations. The Journal Impact Factor (JIF) is the most widely used, in the belief that it may tend to correspond to human judgements of journal quality in some subjects (e.g., Gordon, 1982). Moreover, field normalised indicators, such as the Mean Normalised Citation Score (MNCS) (Waltman, van Eck, van Leeuwen, Visser, & van Raan, 2011a,b) are commonly used for evaluations and reports (e.g., EC, 2007; Elsevier, 2013; NIFU, 2014; NSF, 2014). Most citation indicators do not cope well with the highly skewed nature of sets of citation counts, however, and this has led to a shift toward percentile-based indicators, such as the proportion of a department's outputs in the most highly cited 10% in its field. This approach has been criticised in turn on the basis that percentile-based indicators are imprecise and that geometric mean field normalised citation counts is more precise and therefore more likely to detect genuine differences and less likely to indicate spurious differences (Thelwall, 2016; for a similar JIF argument see also: Zitt, 2012). This conclusion was drawn based upon a set of limiting assumptions, however, and is not therefore definitive. In order to give a more conclusive statement about which indicator is the most suitable for comparing the impacts of sets of articles, theoretical insights are needed into the mathematical properties of sets of citation counts so that randomness within citation counts can be effectively separated from any underlying patterns. This is particularly important because in some cases the decision about which indicator to use affects importance of a set of articles, not because of imprecision in the indicators but because different indicators can give fundamentally different answers (Thelwall, 2016). For example, an indicator to compare the citation impact of nations (e.g., Albarrán, Perianes-Rodríguez, & Ruiz-Castillo, 2015) would not be useful if it could be shown to be an imprecise side-effect of a more fundamental and more precisely measurable property of the citation distribution. In some cases the choice of indicator may be driven by policy decisions but even these need to be informed by an understanding about the properties of the different indicators.

Two attractive distributions for citation data are the power law (Clauset, Shalizi, & Newman, 2009; Egghe, 2005), with probability density function $f(x) = Ax^{-\alpha}$ for $x > c > 0$ (for some c>0, and where A is chosen to ensure that the probabilities sum to 1, so the distribution has only one free parameter, α) and the Yule process (Brzezinski, 2015), with its limiting equilibrium form having probability mass function $f(x) = \rho B(x, \rho + 1)$ for $x = 1, 2, \ldots$, where B is the Euler beta function. Both can be generated by cumulative advantage processes in which the probability that an article attracts citations is related to the number of citations that it already has, so that highly cited articles naturally attract an increasingly large share of new citations (de Solla Price, 1976). This is intuitively reasonable because articles can be found through citations from other articles and by searches in digital libraries that use citation counts within their search results ranking mechanisms (Lawrence, Giles, & Bollacker, 1999). Both of these distributions fit citation data well asymptotically (Brzezinski, 2015) but poorly overall (Thelwall & Wilson, 2014a, in press). The poor overall fit is due to both having monotonic decreasing (point mass) functions, whereas many subject areas have citation count modes of at least 1 (Thelwall, 2016), indicating that they are increasing for some of their values. These distributions only model positive values but can model the full range of citation counts if 1 is added first. This could be thought of as adding a citation to all papers to reflect an implicit self-citation or the basic value that an article has just from being published (see also: de Solla Price, 1976).

Several alternative distributions have been proposed to solve the problem of the ability of pure cumulative advantage distributions to account for the full range of citation counts. The hooked power law (Pennock, Flake, Lawrence, Glover, & Giles, 2002), with probability mass function $f(x) = A(B + x)^{-\alpha}$ (where A is chosen to ensure that the probabilities sum to 1, so the distribution has two free parameters, B and α). It is a discrete version of the Pareto type II distribution (Burrell, 2008) or, more precisely, the Lomax distribution (Lomax, 1954). The Lomax distribution is a Pareto type I distribution shifted to start at zero rather than 1. In the continuous case (i.e., the Lomax distribution) the additional parameter $B$ is a scale parameter (larger values indicate a more spread out distribution) whilst α is a shape parameter. It asymptotically converges to a power law and fits sets of citation counts substantially better than the power law (Eom & Fortunato, 2011; Thelwall & Wilson, 2014a, in press).

The lognormal distribution $ln\mathcal{N}(\mu, \sigma^2)$ (Limpert, Stahel, & Abbt, 2001) with probability density function $f(x) = \frac{1}{x\sigma\sqrt{2\pi}} e^{-\frac{(\ln(x)-\mu)^2}{2\sigma^2}}$ has a fundamentally different basis than the previous three distributions because it is not based on cumulative advantage processes, but is based on the related idea of data being multiplicative rather than additive. (Limpert, Stahel, & Abbt, 2001). More specifically, a continuous lognormal distribution can arise in the limit when positive, independent identically distributed random variables are multiplied together, although it is not clear how this relates to citation distributions. Like the hooked power law, it has two parameters. Its location and scale parameters are the mean and standard deviation, respectively, of the distribution of the natural logarithm of the variable. In its discretised version, the probability of a discrete value $x$ is the integral of the above lognormal probability density function in the unit interval around $x$, $\frac{1}{B}\int_{x-0.5}^{x+0.5} f(x)dx$. Here the constant $B = \int_{0.5}^{\infty} f(x)dx$ compensates for the interval (0,0.5] that is not used for any integer value of $x$. This arbitrary removal of the interval (0,0.5] is aesthetically undesirable and it may be that alternative discretisation solutions that are almost equivalent in practice could be found that do not involve removing any of the original distribution. The location and scale terminology is used here for the corresponding discretised lognormal parameters, even though these do not fit the classical definitions of location and scale after discretisation. The discretised lognormal distribution will be denoted $l\ddot{n}\mathcal{N}(\mu, \sigma^2)$ and fits sets of citation counts from a single field and year much better than the power law and about as well as the hooked power law, depending on the particular field and year (Eom & Fortunato, 2011; Thelwall & Wilson, 2014a, in press). It also fits the distribution of citation counts for articles from a single academic journal and year, for almost all Web of Science journals (Stringer, Sales-Pardo, & Amaral, 2010). Overall, the discretised lognormal distribution fits citation counts (with one added) for individual years less well than does the hooked power law for at least two thirds of Scopus fields (Thelwall, in press).

Although following standard practice in the literature, the discretisation of the lognormal distribution uses integration, the discretisation for the hooked power law uses an alternative approach of using the probability density function as a point mass function, with a correction constant. This approach was used for convenience of calculation but it seems likely these strategies have little difference in practice and do not affect the results.

There has been a claim that citation counts for sets of articles from a single subject and year may follow a discretised lognormal distribution with a variable location parameter but with a common scale parameter $\sigma = \sqrt{1.3} \cong 1.14$ (Evans, Kaube, & Hopkins, 2012; Radicchi, Fortunato, & Castellano, 2008; see also: Perianes-Rodríguez, & Ruiz-Castillo, in press). This is not true for all fields, as can be shown by an analysis of related percentile indicators (Waltman, van Eck, & van Raan, 2012). It is also undermined by datasets with scale parameters varying from 1 to 1.3 (Radicchi, Fortunato, & Castellano, 2008) and from 1.0 to 1.67 (Thelwall, 2016), although no previous studies have tested for statistically significant differences in scale parameters between fields.

The negative binomial distribution (Hilbe, 2011) is another natural choice for citation counts because it is integer valued and highly skewed but it does not fit as well as the shifted power law and discretised lognormal distribution (Low, Wilson, Thelwall, 2015; Thelwall & Wilson, 2014b; Thelwall, in press) because it cannot accommodate the non-trivial amount of very highly cited articles found in typical sets of articles. Stopped sum models (Neyman, 1939) have also been proposed and these seem to fit citation data slightly better than the discretised lognormal and hooked power law overall (Low, Wilson, Thelwall, 2015), but are complex to analyse and are unstable in terms of their parameters and so are not good choices for analyses.

Other distributions that fit sets of citation counts to some extent include a double exponential-Poisson law (Vieira & Gomes, 2010) but this seems overly complex and has only been tested for four fields. An alternative approach is to study the transformations necessary to generate a standard distribution for citation counts (Radicchi, & Castellano, 2012). The stretched exponential function and the q-exponential distribution (which extends the Lomax distribution by allowing different support intervals) have also been shown to fit very broad areas of scholarship reasonably well, such as all natural sciences and engineering as a single category (Wallace, Larivière, & Gingras, 2009). The q-exponential also seems to fit citation distributions for entire countries well (Anastasiadis, de Albuquerque, de Albuquerque, & Mussi, 2009).

Of the two leading choices for modelling the full range of citation counts, the hooked power law and the discretised lognormal distribution, the second is the most useful for theoretical investigations of indicators because its properties are relatively well known and its two parameters separate out the mean ($e^{\mu+\sigma^2/2}$ in the continuous case) and standard deviation ($\sqrt{(e^{\sigma^2}-1)e^{2\mu+\sigma^2}}$ in the continuous case) to a large extent for the typical parameters found within citation distributions. Separating out the mean and standard deviation is important for differentiating between indicators that are affected differently by the standard deviation.

Although previous studies of citation distributions have compared different models to assess which is the best match for a particular set of citation counts, none have directly assessed the plausibility of any given distribution. In other words, none have directly tested whether it would be *reasonable* to believe that set of citation counts had been derived from a specific statistical distribution. It is also possible the goodness of fit of a distribution and the parameters derived from it could be affected by imperfect article subject categorisations, by subfields with different citation norms and by interdisciplinary research. For instance, suppose that a subject containing sub-fields with different rates of attracting citations is analysed. Then even though each subfield may fit one particular distribution well and may even have a common scale parameter in the case of the discretised lognormal distribution (Radicchi, Fortunato, & Castellano, 2008), the overall fit may be poor and the common scale parameter may be affected by differing subfield location parameters. This article addresses these two issues and discusses their combined implications for the choice of field-normalised citation indicator.

## 2. Research questions

The goal of this paper is to assess the extent to which the discretised lognormal and hooked power law distributions could be regarded as reasonable models for citation counts in all academic subjects and years. As discussed above, this is important because some practical issues, such as the best choice of citation indicator, have theoretical dimensions that knowledge of the distribution of citation counts can help to resolve. For example, the geometric mean has been argued to be superior to the arithmetic mean for citation counts and this can be checked for specific distributions (Thelwall, 2016). The first two research questions address whether the two distributions are plausible fit for sets of citation counts. The term *plausible* here means that distribution fitting tests would not reject the null hypothesis that a model fits the data.

1. Are the discretised lognormal and hooked power law distributions the right shape for the set of citation counts for a single subject or year?
2. Are the discretised lognormal and hooked power law distributions plausible candidates as an exact fit for the set of citation counts for a single subject or year?

Whether or not a model is a plausible fit to a data set, it may still be used in practice as the best available approximation. It would sometimes also be convenient to be able to model all citation data with the lognormal distribution even if the hooked power law was a better fit. This is because the parameters of the lognormal distribution are more stable and more easily interpreted. The next two questions investigate the extent to which the two distributions are similar enough to be interchangeable, in the sense that it would be unsurprising if one fitted a set of citation counts that better matched the other. This is important both because previous studies have found the best fitting distribution to depend on the subject area and if they were essentially interchangeable then this would support a conclusion that a single distribution might be appropriate for all subjects.

3. Are the discretised lognormal and hooked power law distributions interchangeable for data sets of citation counts to articles from an individual academic subject?
4. Are the discretised lognormal and hooked power law distributions interchangeable for pure citation-like distributions?

The final question addresses whether the scale parameter of the discretised lognormal distribution could be constant across all homogenous sets of articles (as argued in: Radicchi, Fortunato, & Castellano, 2008). The other parameters are not investigated because there has been no claim that they might be constant, but the results should give insights into how parameters in general might be affected by combining distributions. This extends the theme from the previous two research questions of investigating the extent to which there are alternative distributions or parameters to those that give the optimal fit for a particular data set.

5. Could sets of citation counts from a single subject and year always have the same scale parameter when a discretised lognormal distribution is fitted?

# 3. Data

This article analyses simulated, real, and bootstrapped real data using confidence intervals and the Vuong test. This section details the data sets and the following section explains which research questions each one was used for. The research questions or other relevant sections are also listed in square brackets [>>] after each data set description.

## 3.1 Real data sets

Scopus was chosen as the source of the citation data. Although Web of Science data would probably have given similar results (e.g., Archambault, Campbell, Gingras, & Larivière, 2009), Scopus has wider coverage (López-Illescas, de Moya-Anegón, & Moed, 2008; Moed, & Visser, 2008) and so was selected instead.

**Articles from 2006**. For the real data, 23 Scopus subcategories were chosen and up to 10,000 citation counts for journal articles were downloaded for each one for the year 2006, excluding reviews in order to keep the data as homogeneous as possible. The year 2006 was selected to balance the needs for recent data and large enough citation counts for a powerful statistical analyses. For subcategories having more than 10000 articles, the first 5000 and last 5000 in each year were downloaded, giving a balanced set. In these cases the full sets of articles could not be downloaded because there is a limit of 5000 articles per query.

The 23 Scopus subcategories were systematically selected in order to give a wide range of different fields. For this, the 6th subcategory of each broad Scopus category was chosen except that subcategories with less than 500 articles were rejected as too small and replaced with an alternative. Subcategories were chosen in order to give a relatively narrow disciplinary focus that would give a better chance of producing a pure distribution than broad categories. The choice of the 6th subcategory was arbitrary, constrained only by a desire to avoid subcategories that had been analysed in other published research. When needed, alternative subcategories were chosen by taking the 7th, or subsequent subcategories, as necessary, and cycling back to the 2nd subcategory after reaching the end and if there were originally less than 6 subcategories of a broad category. The 1st subcategory was always ignored because this is a miscellaneous collection and is therefore presumably less homogenous in subject focus than the other areas. Some broad categories were excluded because all subcategories had less than 500 journal articles in 2006. An extra subcategory, the 16th, Cultural Studies, was selected from the broad category with the most subcategories, Social Sciences, because its breadth justified an extra contribution. The full range of subcategories chosen was: Food Science; Cancer Research; Marketing; Filtration and Separation; Physical and Theoretical Chemistry; Computer Science Applications; Management Science and Operations Research; Geochemistry and Petrology; Economics and Econometrics; Energy Engineering and Power Technology; Computational Mechanics; Global and Planetary Change; Virology; Metals and Alloys; Control and Optimization; Critical Care and Intensive Care Medicine; Developmental Neuroscience; Pharmaceutical Science; Nuclear and High Energy Physics; Neuropsychology and Physiological Psychology; Health social science; Cultural Studies; Health Information Management. The total number of article citation counts downloaded was 135,970. [>>**RQ1,2,5**]

**Articles from 2005 to 2014**. In order to investigate changes over time within a subject, Scopus citation counts for all journal articles in five of the subcategories were downloaded for each year from 2005 to 2014: Filtration and Separation; Energy Engineering and Power Technology; Control and Optimization; Nuclear and High Energy Physics; Health social science. These subcategories were selected systematically from the initial set of 23 using a randomly generated starting point. The total number of article citation counts downloaded was 306,124. [>>**Discussion section**]

## 3.2 Bootstrapped samples

For some of the statistical tests bootstrapped samples were needed to assess the variability of the parameters or distribution fit statistics. For each of the main set of 23 data sets, bootstrapped data sets were generated by sampling with replacement from the real citation count data. For example, if a data set contained the citation counts {2,3} then a bootstrapped sample with size 2 from these could be {2,2}, {2,3} or {3,3}. Sampling with replacement ensures that the bootstrapped samples are extremely unlikely to ever be identical to each other. The sampling with replacement was conducted using a random number generator in the statistical package R.

**Same size bootstrapped data set**. The same size bootstrapped data sets contained an identical number of articles to the original subject. For instance, the real data from Marketing contained 2260 articles so the same size bootstrapped samples for Marketing also contained 2260 articles. The bootstrapping was repeated 1000 times, generating 1000x23 bootstrapped samples, one for each subject, containing a total of 1000x135,970 = 135,970,000 bootstrapped citation counts. [>>**RQ3,5**]

**Size 500 bootstrapped data set**. A sample of 500 articles was randomly drawn from each subject to allow direct comparisons between subjects of unequal sizes. For the experiments, bootstrapping was applied 1000 times, generating 1000x23 bootstrapped samples, one for each subject, containing a total of 1000x23x500 = 11,500,000 bootstrapped citation counts. [>>**RQ1,2,5**]

## 3.3 Simulated samples

Some tests needed simulated pure data derived from the discretised lognormal and hooked power law distributions.

**Same size simulated discretised lognormal data sets**. For each of the 23 subjects, a discretised lognormal distribution was modelled with the same scale and location parameters as the medians of those fitted to the bootstrapped citation count data for the subjects. A simulated data set of the same size as the original data set for the subject was then generated from this using the *dist_rand* function in the R powerRlaw package (Gillespie, 2013, 2015). The simulation was repeated 1000 times, generating 1000x23 simulated discretised lognormal samples, one for each subject, containing a total of 1000x135,970 = 135,970,000 simulated citation counts. [>>**RQ4**]

**Size 500 simulated discretised lognormal data sets**. The discretised lognormal simulation was used again with a fixed sample size of 500 articles to allow direct comparisons between subjects. The simulations were repeated 1000 times for each subject, generating 1000x23 simulated samples, one for each subject, containing a total of 1000x23x500 = 11,500,000 simulated citation counts. [>>**RQ4**]

**Same size simulated hooked power law data sets**. For each of the 23 subjects, a hooked power law distribution was set up with the same α and B parameters as the medians of those fitted to the bootstrapped citation count data for the subjects. A simulated data set of the same size as the original data set for the subject was generated from this using the *runif* function in the R stats package to generate random numbers and specially written code to convert them into random integers from the hooked power law distribution[1]. The simulation was repeated 1000 times for each subject, generating 1000x23 simulated hooked power law samples, one for each subject, containing a total of 1000x135,970 = 135,970,000 simulated citation counts. [>>**RQ4**]

**Size 500 simulated hooked power law data sets**. The hooked power law simulations was also generated with a fixed sample size of 500 articles to allow direct comparisons between subjects. The simulation was repeated 1000 times, generating 1000x23 simulated samples, one for each subject, containing a total of 1000x23x500 = 11,500,000 simulated citation counts. [>>**RQ4**]

## 4. Methods

### 4.1 Methods for RQ1: Shape of the discretised lognormal and hooked power law distributions

To compare the shape of the fitted distributions with the data sets, each distribution was first fitted to each real data set (articles from 2006: both the full version and a bootstrapped size 500 version) using the maximum likelihood principle in the R poweRlaw package (Gillespie, 2013, 2015) and the optimal location and scale parameters were recorded. The hooked power law, which was fitted using the maximum likelihood principle with custom-written R code (see Appendix).

A comparison of the theoretical distribution of citation counts for the best fitting discretised lognormal and hooked power law with the empirical distribution can shed light on any systematic points on which the data deviates from the models. The standard method for this is to compare the theoretical and empirical cumulative distributions using a Q-Q plot but Q-Q plots are an imperfect tool for the lognormal distribution (Das, & Resnick, 2008). Direct comparisons of the cumulative distributions are simpler to interpret and are especially helpful for discrete distributions because individual exact values are particularly important. Such plots were created for each of the 23 subjects, one for the hooked power law and one for the discretised lognormal, and were systematically compared for the value at zero, at the empirical median, and at the upper end by the author. Whilst subjective visual interpretations are perhaps not optimal for quantitative analyses, they are standard for analysing distribution fits in statistics, such as with Q-Q plots. These subjective comparisons can be cross-checked using the figures in the online supplement.

---

[1] https://dl.dropboxusercontent.com/u/14001543/hook%20lognormal%20plausible%20paper%20info.zip

## 4.2 Methods for RQ2: Plausibility of the discretised lognormal and hooked power law distributions

The Kolmogorov-Smirnov test (Massey, 1951; Pettitt & Stephens, 1977) was used to check whether it is plausible that real citation data derives from a discretised lognormal distribution. The Kolmogorov-Smirnov statistic is essentially the maximum distance between the theoretical cumulative distribution function and the cumulative distribution of the data. Although there are alternatives that are more powerful in some respects, such as the Cramér-von Mises criterion (Choulakian, Lockhart, & Stephens, 1994), which is based on the square of the distance between the data and the theoretical distribution, and the Anderson Darling test (Anderson & Darling, 1954), which is a version of Cramér-von Mises that weights the tail of the distribution more heavily, none are universally the most powerful and the Kolmogorov-Smirnov test has the advantage of simplicity.

The Kolmogorov-Smirnov test can be used to test the hypothesis that data is drawn from a continuous theoretical distribution by comparing the Kolmogorov-Smirnov statistic from a data set with tabulated standard values for continuous distributions. For discrete distributions, there is a conservative equivalent procedure but a less conservative approach, used here, is to compare the value with the Kolmogorov-Smirnov statistics of (say) 1000 random samples drawn from the same distribution to estimate the probability that the Kolmogorov-Smirnov statistics for the real data could have been derived from a random sample from the distribution (Arnold, & Emerson, 2011; Goodman, 1954).

To test whether each set of simulated citation data could *plausibly* be derived from a discretised lognormal distribution, the distribution was again fitted to each real data set (articles from 2006: both the full version and a bootstrapped size 500 version) using the maximum likelihood principle in the R poweRlaw package and the optimal location and scale parameters were recorded. The null hypothesis that the data was drawn from the discretised lognormal distribution with the fitted location and scale parameters was tested using the R discrete goodness of fit testing package dgof (Arnold, & Emerson, 2011), which estimates this probability through simulated sampling from the same distribution. The same tests were conducted for the hooked power law, which was fitted using the maximum likelihood principle with custom-written R code (Thelwall & Wilson, 2014a; see the Appendix). The code did not always converge because for large parameter values changes in the α and B values can almost cancel each other out (Thelwall & Wilson, 2014a). As can be seen from the standard formulae for the mean and standard deviation of the Lomax distribution, this is because the ratio of the two is approximately proportional to both the mean and standard deviation when they are both large. In practice, large increases in these parameters are needed to make minor changes in the shape of the distribution, preventing convergence.

The above test was repeated for the size 500 bootstrapped data set to allow fair comparisons between subjects.

### 4.3 Methods for RQ3: Interchangeability of the discretised lognormal and hooked power law distributions for subjects

To assess whether the discretised lognormal and hooked power law are interchangeable for sets of citation counts from individual subjects and years, both were fitted to the two bootstrapped data sets and the fit compared using the Vuong test (Vuong, 1989; Wilson, 2015; see also: Akaike, 1974). The Vuong test generates a z value that is approximately normally distributed and hence can be used to assess whether one distribution fits data statistically significantly better than another[2]. The outcomes of the 1000 Vuong tests for each subject were tallied in order to assess how often one test was a statistically significantly better fit than the other. The 1000 bootstrapped samples were also used to generate 95% confidence intervals for the Vuong z value for comparison purposes.

### 4.4 Methods for RQ4: Interchangeability of the discretised lognormal and hooked power law for citation-like theoretical distributions

To assess whether the discretised lognormal and hooked power law are interchangeable for pure simulated citation counts from either the hooked power law or the discretised lognormal distribution, both were fitted to each sample in the first four simulated data sets described above and their fits compared using the Vuong test as above.

### 4.5 Methods for RQ5: Constant standard deviation tests

To assess whether the scale parameters of the discretised lognormal distribution fitted to the real data could conceivably be the same, 95% confidence intervals for the scale parameter of each distribution were generated using the 1000 size 500 bootstrapped data sets. As above, the confidence interval limits were derived from the 25th smallest and 25th largest of the 1000 scale parameters for each subject (for confidence interval issues with the lognormal distribution, see: Zhou & Gao, 1997).

## 5. Results

### 5.1 RQ1: Are the discretised lognormal and hooked power law distributions the right shape for homogeneous citation count distributions?

As discussed in the methods, for each of the 23 subjects, the hooked power law and discretised lognormal were compared with the empirical data and the results are summarised in Table 1. For the lognormal distribution, the empirical cumulative distribution was always higher for uncited articles (1 in the data because of adding 1 before fitting the distribution), and lower for the middle (median) of the distribution. With three exceptions, the empirical data was also higher for the top of the distribution, meaning that the empirical data reached the maximum before the theoretical distribution did. There was a tendency for the same pattern to occur for the hooked power law, but it was much weaker. Figure 1 shows the first (in Scopus order) subcategory, Food Science and figures 2 and 3 illustrate the comparison for this subcategory. The other figures, including figures for bootstrapped samples of size 500 for all subjects, are available online (see Appendix). Overall, this gives strong evidence that the discretised lognormal distribution has the wrong shape for citation counts and suggests that the hooked power law may also have the wrong shape.

---

[2] https://dl.dropboxusercontent.com/u/14001543/hook%20lognormal%20plausible%20paper%20info.zip

**Table 1**. A comparison between the empirical cumulative distribution and the discretised lognormal or hooked power law distributions fitting best for each of the 23 subjects.

| Empirical data* | Ln bottom | Ln middle | Ln top | Hook bottom | Hook middle | Hook top |
|---|---|---|---|---|---|---|
| Cancer Research | + | - | + | + | - | = |
| Computational Mechanics | + | - | + | = | = | = |
| Computer Science Applications | + | - | + | + | - | = |
| Control and Optimization | + | - | = | - | = | = |
| Critical Care & Intensive Care Medicine | + | - | + | + | - | + |
| Cultural Studies | + | - | = | + | - | + |
| Developmental Neuroscience | + | - | + | - | = | = |
| Economics and Econometrics | + | - | + | + | = | = |
| Energy Engineering & Power Tech. | + | - | + | + | - | + |
| Filtration and Separation | + | - | + | + | = | = |
| Food Science | + | - | + | + | - | = |
| Geochemistry and Petrology | + | - | + | - | = | = |
| Global and Planetary Change | + | - | + | + | - | = |
| Health  social science | + | - | + | + | - | = |
| Health Information Management | + | - | = | + | - | + |
| Management Sci. & Operations Res. | + | - | + | = | = | = |
| Marketing | + | - | + | + | = | = |
| Metals and Alloys | + | - | + | + | - | + |
| Neuropsych. & Physiological Psych. | + | - | + | + | - | = |
| Nuclear and High Energy Physics | + | - | + | + | - | = |
| Pharmaceutical Science | + | - | + | + | - | + |
| Physical and Theoretical Chemistry | + | - | + | = | = | = |
| Virology | + | - | + | = | - | = |
| **Higher total** | 23 | 0 | 20 | 16 | 0 | 6 |
| **Same total** | 0 | 0 | 3 | 4 | 9 | 17 |
| **Lower total** | 0 | 23 | 0 | 3 | 14 | 0 |
| **Overall total** | **23** | **23** | **23** | **23** | **23** | **23** |

* +: empirical data higher than the theoretical value; =: empirical data about the same as the theoretical value; -: empirical data lower than the theoretical value

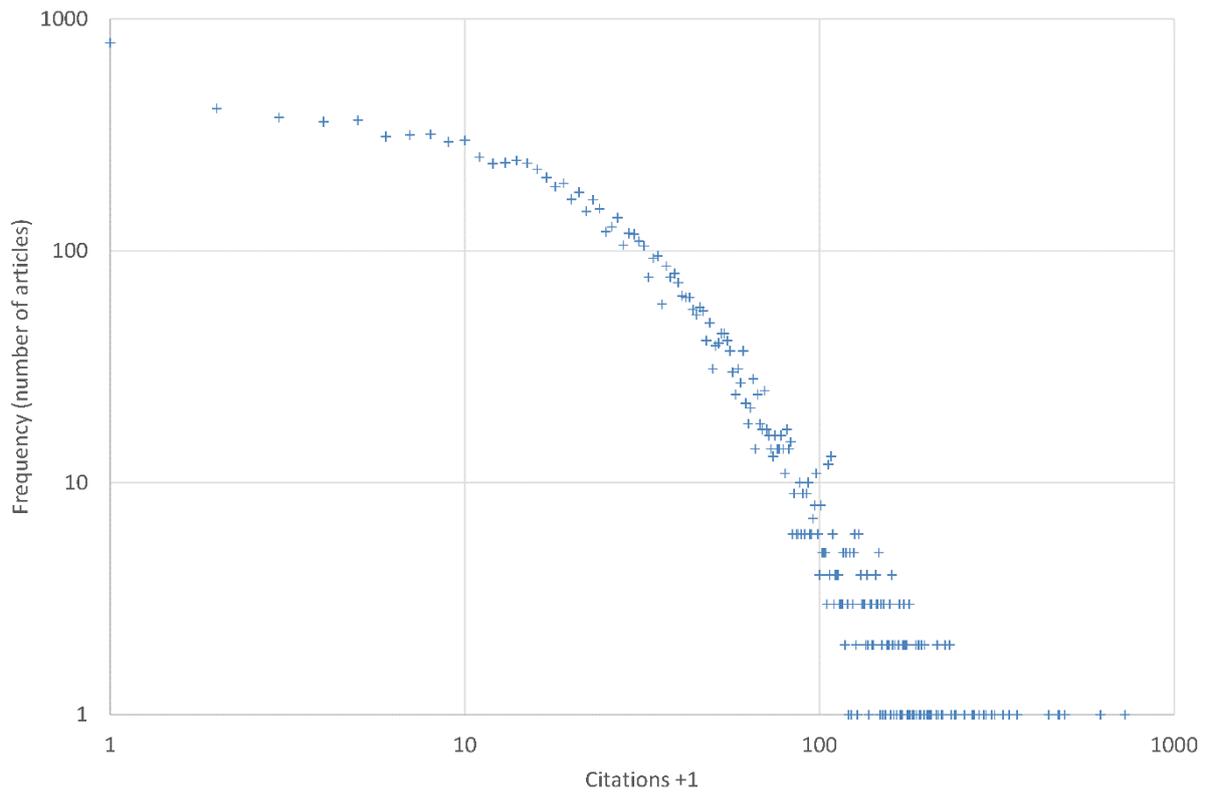

**Figure 1**. The distribution of citation counts for Food Science, after adding 1. Note the logarithmic scales.

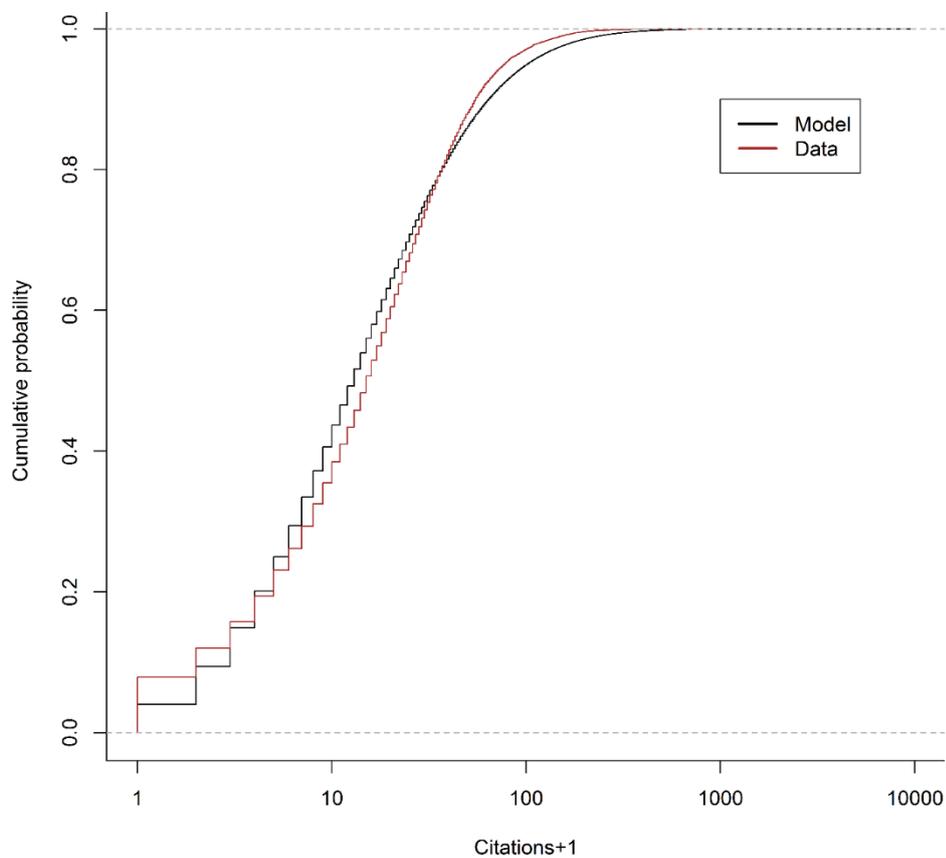

**Figure 2**. The empirical and theoretical discretised lognormal cumulative distributions for Food Science. The empirical data in this case is higher for the bottom and top of the distribution but lower for the middle of the distribution.

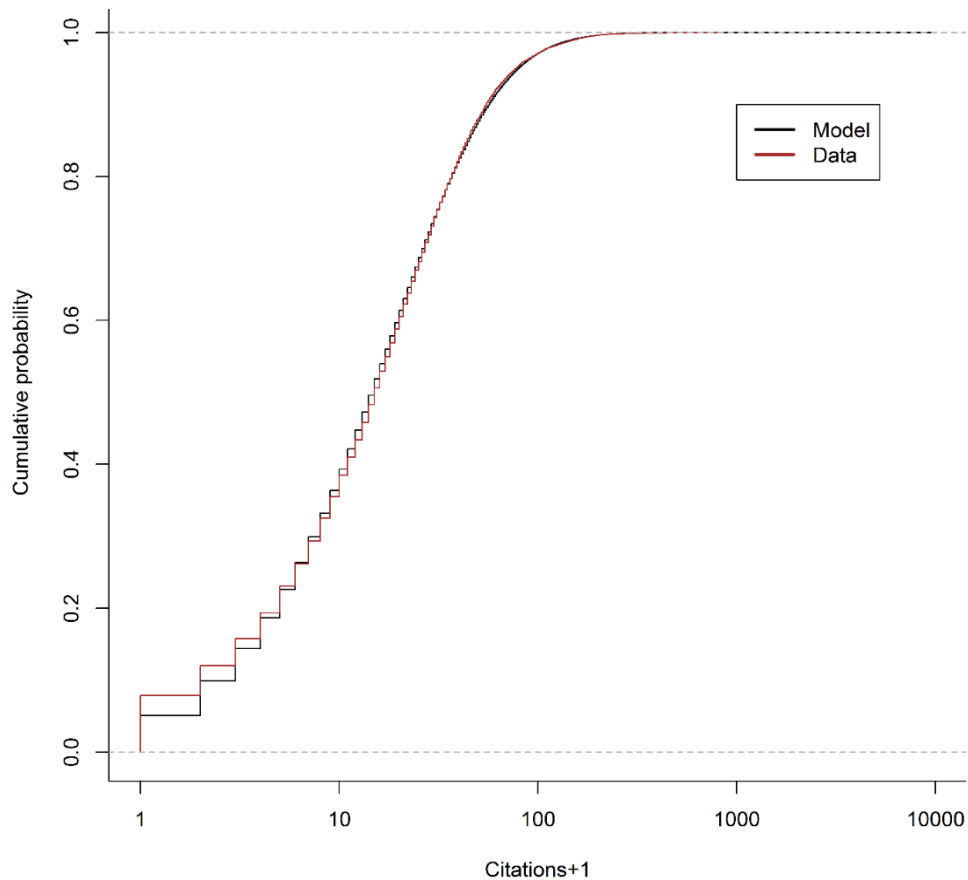

**Figure 3**. The empirical and theoretical hooked power law cumulative distributions for Food Science. The empirical data in this case is higher for the bottom and top of the distribution but lower for the middle of the distribution.

## 5.2 RQ2: Are the discretised lognormal and hooked power law distributions plausible candidates in all subjects for homogeneous citation count distributions?

The lognormal is a *plausible* distribution for 4 out of the 23 subjects according to the Kolmogorov-Smirnov goodness of fit tests and the hooked power law is a *plausible* distribution for 5 subjects (i.e., p values above 0.05 in Table 2). For one subject, both are plausible but for most subjects (16) neither one is plausible. Thus, neither distribution is a candidate for a universal exact distribution for citation counts for Scopus subcategories. Considering that only one of the larger subject areas has a plausible distribution (Computational Mechanics, n=7776), it seems possible that neither distribution is an exact fit for any subject area and year but that they are both close approximations for some subject areas and years.

**Table 2**. Kolmogorov-Smirnov tests of the null hypotheses that the samples are drawn from a discretised lognormal distribution or a hooked power law distribution for the *articles from 2006* data set. See the Appendix for online figures for each subject.

| Subject | N | Ln μ | Ln σ | Ln KS | Ln KS p | Hook α | Hook B | Hook KS | Hook KS p | Plaus-ible |
|---|---|---|---|---|---|---|---|---|---|---|
| Cancer Research | 9994 | 2.77 | 1.4 | 0.067 | 0 | 3.94 | 67.9 | 0.043 | 0 | |
| Computational Mechanics | 7776 | 2.19 | 1.17 | 0.026 | 0 | 4.87 | 46.1 | 0.008 | 0.519 | H |
| Computer Science Applications | 8148 | 2.19 | 1.4 | 0.043 | 0 | 3.11 | 24.3 | 0.033 | 0 | |
| Control and Optimization | 1043 | 2.08 | 1.11 | 0.024 | 0.416 | 5.07 | 41.9 | 0.028 | 0.231 | H,L |
| Critical Care and Intensive Care Medicine | 2625 | 1.75 | 1.82 | 0.065 | 0 | 2.06 | 7.1 | 0.074 | 0 | |
| Cultural Studies | 4848 | -0.38 | 1.73 | 0.01 | 0.395 | 2.26 | 1.3 | 0.021 | 0.005 | L |
| Developmental Neuroscience | 1394 | 2.83 | 1.08 | 0.038 | 0.023 | 11.34 | 258.1 | 0.041 | 0.011 | |
| Economics and Econometrics | 9974 | 2.23 | 1.39 | 0.024 | 0 | 3.08 | 24.4 | 0.022 | 0 | |
| Energy Engineering and Power Technology | 7833 | 1.33 | 1.79 | 0.045 | 0 | 2.07 | 4.7 | 0.069 | 0 | |
| Filtration and Separation | 3282 | 2.18 | 1.39 | 0.048 | 0 | 3.56 | 31.5 | 0.045 | 0 | |
| Food Science | 9992 | 2.54 | 1.26 | 0.059 | 0 | 5.76 | 89.8 | 0.028 | 0 | |
| Geochemistry and Petrology | 8292 | 2.79 | 1.12 | 0.034 | 0 | 6.55 | 126.9 | 0.037 | 0 | |
| Global and Planetary Change | 834 | 3 | 1.35 | 0.048 | 0.032 | 3.5 | 67 | 0.02 | 0.813 | H |
| Health  social science | 4352 | 2.15 | 1.5 | 0.083 | 0 | 3.82 | 37.9 | 0.092 | 0 | |
| Health Information Management | 697 | 1.96 | 1.37 | 0.046 | 0.060 | 3.5 | 23.8 | 0.05 | 0.036 | L |
| Management Science & Operations Research | 3993 | 2.45 | 1.24 | 0.03 | 0.001 | 4.08 | 47.6 | 0.008 | 0.863 | H |
| Marketing | 2260 | 2.43 | 1.33 | 0.033 | 0.007 | 3.52 | 37.8 | 0.016 | 0.464 | H |
| Metals and Alloys | 9964 | 1.22 | 1.53 | 0.011 | 0.089 | 2.38 | 5.2 | 0.026 | 0 | L |
| Neuropsychology & Physiological Psychology | 2927 | 2.59 | 1.39 | 0.074 | 0 | 4.55 | 72.6 | 0.051 | 0 | |
| Nuclear and High Energy Physics | 9994 | 2.34 | 1.41 | 0.042 | 0 | 3.23 | 30.8 | 0.032 | 0 | |
| Pharmaceutical Science | 9228 | 1.9 | 1.61 | 0.082 | 0 | 2.95 | 19.4 | 0.098 | 0 | |
| Physical and Theoretical Chemistry | 9986 | 2.46 | 1.18 | 0.032 | 0 | 4.77 | 59.3 | 0.02 | 0 | |
| Virology | 6534 | 2.81 | 1.05 | 0.058 | 0 | 14.74 | 329.5 | 0.058 | 0 | |

To enable comparable results between subject areas, the test was repeated for a bootstrapped random sample of 500 articles from each subject area. From this sample, the lognormal is a plausible distribution for 14 out of the 23 subjects according to the Kolmogorov-Smirnov test and the hooked power law is a plausible distribution for 16 subjects (i.e., p values above 0.05 in Table 3). For 11 subjects, both are plausible and for 4 subjects neither is plausible. The greater number of plausible distributions in this case is due to the lower power of the test for smaller sample sizes. The results confirm that a distribution can seem plausible for smaller sample sizes even though they are not plausible for large sample sizes.

**Table 3**. Kolmogorov-Smirnov tests of the null hypotheses that the samples are drawn from a discretised lognormal distribution or a hooked power law distribution for random samples of size 500 drawn from the *articles from 2006* data set. Since the samples are randomly drawn, other results are possible for each subject and year. See the Appendix for online figures for each subject.

| Subject | N | Ln μ | Ln σ | Ln KS | Ln KS p | Hook α | Hook B | Hook KS | Hook KS p | Plaus-ible |
|---|---|---|---|---|---|---|---|---|---|---|
| Cancer Research | 500 | 2.76 | 1.35 | 0.064 | 0.013 | 4.1 | 69.8 | 0.033 | 0.528 | L,H |
| Computational Mechanics | 500 | 2.24 | 1.13 | 0.038 | 0.316 | 6.02 | 65.1 | 0.024 | 0.807 | L,H |
| Computer Science Applications | 500 | 2.15 | 1.46 | 0.067 | 0.011 | 3.01 | 22.8 | 0.047 | 0.148 | H |
| Control and Optimization | 500 | 2.15 | 1.16 | 0.031 | 0.53 | 4.43 | 37.5 | 0.026 | 0.736 | L,H |
| Critical Care & Intensive Care Med. | 500 | 1.61 | 1.92 | 0.069 | 0.009 | 1.92 | 5.1 | 0.081 | 0.002 | |
| Cultural Studies | 500 | 0.13 | 1.47 | 0.016 | 0.917 | 2.57 | 2.1 | 0.021 | 0.708 | L,H |
| Developmental Neuroscience | 500 | 2.83 | 1.07 | 0.041 | 0.272 | 9.43 | 205.9 | 0.047 | 0.158 | L,H |
| Economics and Econometrics | 500 | 2.25 | 1.37 | 0.041 | 0.253 | 3.4 | 30.5 | 0.028 | 0.704 | L,H |
| Energy Engineering & Power Tech. | 500 | 1.52 | 1.69 | 0.056 | 0.056 | 2.24 | 6.8 | 0.07 | 0.007 | L |
| Filtration & Separation | 500 | 2.21 | 1.4 | 0.066 | 0.015 | 3.65 | 34.6 | 0.054 | 0.067 | H |
| Food Science | 500 | 2.55 | 1.19 | 0.077 | 0.002 | 7.59 | 125.7 | 0.04 | 0.302 | H |
| Geochemistry and Petrology | 500 | 2.84 | 1.02 | 0.031 | 0.612 | 7.84 | 160.8 | 0.074 | 0.005 | L |
| Global & Planetary Change | 500 | 3.03 | 1.39 | 0.058 | 0.043 | 3.27 | 62.2 | 0.028 | 0.753 | H |
| Health  social science | 500 | 2.01 | 1.55 | 0.065 | 0.01 | 3.37 | 27 | 0.095 | 0.001 | |
| Health Information Management | 500 | 1.95 | 1.37 | 0.055 | 0.052 | 3.75 | 26.9 | 0.058 | 0.032 | L |
| Management Sci. & Operations Res. | 500 | 2.57 | 1.17 | 0.027 | 0.713 | 4.82 | 67.5 | 0.026 | 0.771 | L,H |
| Marketing | 500 | 2.43 | 1.37 | 0.043 | 0.225 | 3.5 | 38.4 | 0.028 | 0.67 | L,H |
| Metals and Alloys | 500 | 1.16 | 1.52 | 0.015 | 0.981 | 2.42 | 5.1 | 0.031 | 0.483 | L,H |
| Neuropsych. & Physiological Psych. | 500 | 2.64 | 1.36 | 0.063 | 0.026 | 5.02 | 86.6 | 0.043 | 0.242 | H |
| Nuclear & High Energy Physics | 500 | 2.28 | 1.46 | 0.043 | 0.235 | 2.9 | 23.9 | 0.038 | 0.328 | L,H |
| Pharmaceutical Science | 500 | 2 | 1.58 | 0.096 | 0 | 3.64 | 32.3 | 0.112 | 0 | |
| Physical & Theoretical Chemistry | 500 | 2.4 | 1.23 | 0.045 | 0.207 | 4.82 | 58.7 | 0.018 | 0.97 | L,H |
| Virology | 500 | 2.82 | 1.01 | 0.066 | 0.019 | 30.17 | 713.6 | 0.08 | 0.002 | |

Note that if the parameter values given in tables 2 and 3 are used to estimate the arithmetic means of the data using the formulae for the continuous distributions ($e^{\mu+\sigma^2/2}$ for the continuous lognormal and $B/(\alpha-1)$ for the hooked power law) then these give widely differing estimates. This is surprising given that, intuitively, they should be reasonable estimates even if they are only approximations in the discrete case. For example, the average estimate for the mean from the lognormal parameters in Table 2 using the above formula is 25.4 whereas the average estimate for the mean from the hooked power law parameters using the above formula is 14.2. The average mean for the data (after adding 1) differs from both at 22.9, although it is closer to the lognormal estimate. Thus, the optimal parameters for each distribution from a maximum likelihood perspective generate distributions with arithmetic means that are higher (lognormal) or lower (hooked power law) than the data. This is possible due to the highly skewed nature of the data sets.

## 5.3 RQ3: Are the discretised lognormal and hooked power law distributions interchangeable for sets of citation counts from academic subjects?

The hooked power law fits better than, or as well as, the discretised lognormal in all of the bootstrapped samples for 17 subjects (Table 4). Similarly the discretised lognormal fits better than, or as well as, the hooked power law in all of the bootstrapped samples for 5 subjects. In the remaining subject, Control and Optimisation, there was insufficient statistical evidence to suggest that one fits better than the other in 950 of the 1000 cases. This gives robust evidence that one distribution fits better than another for most subjects.

Comparing the results to those in Table 2, the results are not contradictory except in the case of Health Information Management. For this subject, only the lognormal is plausible (Table 2) but the hooked power law tends to be a significantly better fit (Table 4). This is partly due to the minor differences around the significance level in Table 2 but also shows that the two tests are assessing different (but related) aspects of the distribution fits. For some subject categories (e.g., Cancer Research; Critical Care and Intensive Care Medicine), although distribution one is always a significantly better fit than another in the bootstrapped case, it is sometimes the case that neither is plausible (in Table 2). Thus, one distribution can be a robustly better fit than the other even when it is demonstrably not a perfect fit. Presumably in such cases the better fitting distribution has a more closely related shape to that of the empirical data. The unusual case of Control and Optimisation in that both distributions are plausible (Table 2) and neither tends to fit better than the other (Table 4) can only be partly attributed to the lower number of data points (see the results for all subjects for the size 500 data sets below) and so suggests that the data is a reasonably close approximation to both distributions. For example, if it were to be intermediate between the two in some sense then this would explain the results.

**Table 4**. Vuong test z value and 95% confidence intervals from 1000 bootstrapped samples for each subject. The last two columns record the number of tests for which one distribution is a statistically significantly better fit than the other at the $p = 0.05$ level.

| Subject | Sample size (articles) | Lower 95% limit | Median Vuong | Upper 95% limit | Hooked a sig. better fit | Discr. Logn. a sig. better fit |
|---|---|---|---|---|---|---|
| Cancer Research | 9994 | 16.67 | 17.93 | 18.95 | 1000 | 0 |
| Computational Mechanics | 7776 | 2.32 | 4.3 | 5.98 | 989 | 0 |
| Computer Science Applications | 8148 | 4.62 | 6.38 | 7.99 | 1000 | 0 |
| Control and Optimization | 1043 | -1.99 | -0.09 | 1.95 | 22 | 28 |
| Critical Care and Intensive Care Medicine | 2625 | -13.78 | -11.02 | -8.2 | 0 | 1000 |
| Cultural Studies | 4848 | -6.19 | -4.64 | -3.04 | 0 | 999 |
| Developmental Neuroscience | 1394 | 0.19 | NA* | 4.21 | 601 | 0 |
| Economics and Econometrics | 9974 | 2.74 | 4.45 | 6.04 | 998 | 0 |
| Energy Engineering and Power Technology | 7833 | -23.9 | -21.49 | -19.02 | 0 | 1000 |
| Filtration and Separation | 3282 | 1.6 | 3.1 | 4.57 | 916 | 0 |
| Food Science | 9992 | 16.62 | 17.87 | 18.98 | 1000 | 0 |
| Geochemistry and Petrology | 8292 | 1.09 | 3.09 | 5.17 | 882 | 0 |
| Global and Planetary Change | 834 | 1.27 | 2.97 | 4.27 | 894 | 0 |
| Health  social science | 4352 | 2.68 | 4.53 | 6.45 | 998 | 0 |
| Health Information Management | 697 | 0.69 | 2.38 | 3.53 | 685 | 0 |
| Management Science & Operations Res. | 3993 | 2.45 | 4.46 | 6.04 | 997 | 0 |
| Marketing | 2260 | 2.59 | 4.09 | 5.36 | 996 | 0 |
| Metals and Alloys | 9964 | -13.26 | -11.61 | -10.06 | 0 | 1000 |
| Neuropsychology & Physiological Psychology | 2927 | 7.3 | 8.8 | 10.17 | 1000 | 0 |
| Nuclear and High Energy Physics | 9994 | 4.8 | 6.51 | 8.03 | 1000 | 0 |
| Pharmaceutical Science | 9228 | -5.08 | -2.89 | -0.91 | 0 | 820 |
| Physical and Theoretical Chemistry | 9986 | 3.86 | 5.74 | 7.68 | 1000 | 0 |
| Virology | 6534 | 3.75 | 5.78 | 7.8 | 1000 | 0 |

*The hooked power law did not always fit the bootstrapped samples from this subject (see section 4.2).

Some of the differences between subjects can be due to the smaller sample sizes that make the Vuong test less powerful. To check for this, the tests were repeated but with bootstrapped samples of size 500 rather than the same size as the original sample (Table 5). Although the reduced power of the test is evident in the slightly lower numbers of significant results, this does not affect the overall conclusions.

**Table 5**. Vuong test z value and 95% confidence intervals from 1000 bootstrapped samples of 500 articles for each subject. The last two columns record the number of tests for which one distribution is a statistically significantly better fit than the other at the $p = 0.05$ level.

| Subject | Sample size (articles) | Lower 95% limit | Median Vuong | Upper 95% limit | Hooked a sig. better fit | Discr. Logn. a sig. better fit |
|---|---|---|---|---|---|---|
| Cancer Research | 500 | 2.42 | 3.91 | 5.02 | 992 | 0 |
| Computational Mechanics | 500 | -1.03 | 1.09 | 2.81 | 178 | 0 |
| Computer Science Applications | 500 | -0.24 | 1.52 | 3.04 | 293 | 0 |
| Control and Optimization | 500 | -1.97 | -0.02 | 1.96 | 25 | 25 |
| Critical Care and Intensive Care Medicine | 500 | -7.36 | -4.76 | -2.36 | 0 | 987 |
| Cultural Studies | 500 | -2.95 | -1.45 | 0.2 | 0 | 274 |
| Developmental Neuroscience | 500 | NA* | NA* | 3.32 | 281 | 0 |
| Economics and Econometrics | 500 | -0.79 | 0.95 | 2.38 | 98 | 1 |
| Energy Engineering and Power Technology | 500 | -7.55 | -5.34 | -3.02 | 0 | 997 |
| Filtration and Separation | 500 | -0.6 | 1.13 | 2.51 | 149 | 0 |
| Food Science | 500 | 2.44 | 3.91 | 5.18 | 992 | 0 |
| Geochemistry and Petrology | 500 | -1.13 | 0.83 | 2.8 | 137 | 1 |
| Global and Planetary Change | 500 | 0.37 | 2.2 | 3.53 | 622 | 0 |
| Health  social science | 500 | -0.42 | 1.57 | 3.35 | 317 | 1 |
| Health Information Management | 500 | 0.22 | 1.97 | 3.16 | 504 | 0 |
| Management Science & Operations Res. | 500 | -0.32 | 1.57 | 3.03 | 324 | 0 |
| Marketing | 500 | 0.14 | 1.84 | 3.18 | 446 | 0 |
| Metals and Alloys | 500 | -3.96 | -2.56 | -0.89 | 0 | 782 |
| Neuropsychology & Physiological Psychology | 500 | 2.1 | 3.64 | 4.97 | 983 | 0 |
| Nuclear and High Energy Physics | 500 | -0.16 | 1.42 | 2.86 | 255 | 0 |
| Pharmaceutical Science | 500 | -2.89 | -0.62 | 1.3 | 4 | 106 |
| Physical and Theoretical Chemistry | 500 | -0.66 | 1.35 | 3.13 | 275 | 0 |
| Virology | 500 | NA* | NA* | 3.75 | 354 | 0 |

*The hooked power law did not always fit the bootstrapped samples from this subject (see section 4.2).

## 5.4 RQ4: Are the discretised lognormal and hooked power law distributions interchangeable for theoretical citation-like distributions?

When artificial samples are generated from the discretised lognormal distribution with parameters matching one of the 23 subjects (Table 6), the hooked power law is almost never a better fit (1 case out of 23x1000=23,000). The discretised lognormal tends to be a better fit, as expected, and is a statistically significantly better fit in the vast majority of cases (18,230 out of 23,000 or 79%). The two distributions are statistically indistinguishable (at least by the Vuong test) in the majority of cases for simulated data from four subjects: Filtration and Separation; Global and Planetary Change; Cultural Studies; Health Information Management. This is unexpected given the relatively large sample sizes (from 697 to 4848) and suggests that for some parameter sets, the discretised lognormal is similar to the hooked power law in practice.

When the sample sizes for the simulations are reduced to 500 then the two distributions have similar enough fits to pass the Vuong test a majority of the time (Table 7). This suggests that large samples are needed to show up differences between the distributions.

**Table 6**. Vuong test z value and 95% confidence intervals from 1000 repetitions of sets of simulated citation counts generated from the discretised lognormal distribution. For each subject and each set of 1000 samples, the data was randomly drawn from a discretised lognormal distribution with scale and location parameters that are optimal for the subject. The number of simulated citation counts in each subject is equal to the number of articles in the original data set. The last two columns record the number of tests for which one distribution is a statistically significantly better fit than the other at the $p = 0.05$ level.

| Subject that parameters were taken from | Sample size (articles) | Lower 95% limit | Median Vuong | Upper 95% limit | Hooked a sig. better fit | Discr. Logn. a sig. better fit |
|---|---|---|---|---|---|---|
| Cancer Research | 9994 | -6.37 | -4.39 | -2.38 | 0 | 990 |
| Computational Mechanics | 7776 | -7.93 | -6.08 | -4.23 | 0 | 1000 |
| Computer Science Applications | 8148 | -4.98 | -2.8 | -0.95 | 0 | 791 |
| Control and Optimization | 1043 | -4.35 | -2.58 | -0.7 | 0 | 742 |
| Critical Care and Intensive Care Medicine | 2625 | -5.4 | -3.19 | -1.12 | 0 | 869 |
| Cultural Studies | 4848 | -3.54 | -1.73 | 0.04 | 0 | 399 |
| Developmental Neuroscience | 1394 | -7.04 | -4.96 | -3.1 | 0 | 1000 |
| Economics and Econometrics | 9974 | -5.28 | -3.22 | -1.3 | 0 | 893 |
| Energy Engineering and Power Technology | 7833 | -7.43 | -5.26 | -3.03 | 0 | 999 |
| Filtration and Separation | 3282 | -3.97 | -1.74 | 0.07 | 0 | 422 |
| Food Science | 9992 | -8.08 | -6.08 | -3.97 | 0 | 1000 |
| Geochemistry and Petrology | 8292 | -12.71 | -10.59 | -8.55 | 0 | 1000 |
| Global and Planetary Change | 834 | -4.02 | -1.72 | 0.22 | 0 | 396 |
| Health  social science | 4352 | -4.64 | -2.36 | -0.31 | 0 | 646 |
| Health Information Management | 697 | -3.01 | -0.83 | 0.97 | 1 | 130 |
| Management Science & Operations Res. | 3993 | -5.8 | -3.86 | -2.01 | 0 | 978 |
| Marketing | 2260 | -4.12 | -2.03 | -0.07 | 0 | 521 |
| Metals and Alloys | 9964 | -6.35 | -4.12 | -2.11 | 0 | 980 |
| Neuropsychology & Physiological Psychology | 2927 | -4.22 | -2.13 | -0.14 | 0 | 583 |
| Nuclear and High Energy Physics | 9994 | -5.41 | -3.32 | -1.3 | 0 | 896 |
| Pharmaceutical Science | 9228 | -6.59 | -4.45 | -2.49 | 0 | 995 |
| Physical and Theoretical Chemistry | 9986 | -9.7 | -7.83 | -6.14 | 0 | 1000 |
| Virology | 6534 | -13.71 | -11.69 | -9.68 | 0 | 1000 |

**Table 7**. Vuong test z value and 95% confidence intervals from 1000 repetitions of sets of 500 simulated citation counts generated from the discretised lognormal distribution. For each subject and each set of 1000 samples, the data was randomly drawn from a discretised lognormal distribution with scale and location parameters that are optimal for the subject. The last two columns record the number of tests for which one distribution is a statistically significantly better fit than the other at the $p = 0.05$ level.

| Subject that parameters were taken from | Sample size (articles) | Lower 95% limit | Median Vuong | Upper 95% limit | Hooked a sig. better fit | Discr. Logn. a sig. better fit |
|---|---|---|---|---|---|---|
| Cancer Research | 500 | -3.13 | -0.87 | 0.9 | 2 | 167 |
| Computational Mechanics | 500 | -3.16 | -1.55 | 0.44 | 0 | 334 |
| Computer Science Applications | 500 | -2.89 | -0.74 | 0.96 | 2 | 130 |
| Control and Optimization | 500 | -3.66 | -1.84 | 0.08 | 0 | 452 |
| Critical Care and Intensive Care Medicine | 500 | -3.56 | -1.45 | 0.47 | 0 | 337 |
| Cultural Studies | 500 | -2.43 | -0.61 | 0.98 | 0 | 67 |
| Developmental Neuroscience | 500 | -5.08 | -3.05 | -0.85 | 0 | 832 |
| Economics and Econometrics | 500 | -2.96 | -0.75 | 1.08 | 0 | 119 |
| Energy Engineering and Power Technology | 500 | -3.6 | -1.47 | 0.52 | 0 | 330 |
| Filtration and Separation | 500 | -2.81 | -0.74 | 1 | 0 | 134 |
| Food Science | 500 | -3.28 | -1.36 | 0.55 | 0 | 276 |
| Geochemistry and Petrology | 500 | -4.73 | -2.62 | -0.55 | 0 | 744 |
| Global and Planetary Change | 500 | -3.3 | -1.35 | 0.6 | 0 | 273 |
| Health  social science | 500 | -2.74 | -0.9 | 1.07 | 1 | 140 |
| Health Information Management | 500 | -2.93 | -0.67 | 1.1 | 2 | 121 |
| Management Science & Operations Res. | 500 | -3.26 | -1.36 | 0.49 | 0 | 260 |
| Marketing | 500 | -3.05 | -0.98 | 1.06 | 1 | 175 |
| Metals and Alloys | 500 | -2.93 | -1.07 | 0.94 | 2 | 201 |
| Neuropsychology & Physiological Psychology | 500 | -2.91 | -0.92 | 0.89 | 1 | 144 |
| Nuclear and High Energy Physics | 500 | -2.95 | -0.79 | 1.01 | 0 | 140 |
| Pharmaceutical Science | 500 | -3.1 | -1.19 | 0.89 | 0 | 212 |
| Physical and Theoretical Chemistry | 500 | -3.6 | -1.8 | 0.15 | 0 | 436 |
| Virology | 500 | -5.4 | -3.26 | -1.26 | 0 | 884 |

When artificial samples are generated from the hooked power law distribution with parameters matching one of the 23 subjects (Table 9), the discretised lognormal is almost never a better fit (2 cases out of 23,000). The hooked power law tends to be a better fit, as again expected, and is a statistically significantly better fit in the vast majority of cases (19,818 out of 23,000 or 86%). The two distributions are statistically indistinguishable (at least by the Vuong test) in the majority of cases for three subjects (all of which had the same issue in Table 6): Global and Planetary Change; Cultural Studies; Health Information Management.

When the sample sizes for the simulations are reduced to 500 then the two distributions have similar enough fits to pass the Vuong test a majority of the time (Table 10). This suggests that large samples are needed to show up differences between the distributions.

**Table 9**. Vuong test z value and 95% confidence intervals from 1000 repetitions of sets of simulated citation counts generated from the hooked power law distribution. For each subject and each set of 1000 samples, the data was randomly drawn from a hooked power law distribution with α and B parameters that are optimal for the subject. The number of simulated citation counts in each subject is equal to the number of articles in the original data set. The last two columns record the number of tests for which one distribution is a statistically significantly better fit than the other at the $p = 0.05$ level.

| Subject that parameters were taken from | Sample size (articles) | Lower 95% limit | Median Vuong | Upper 95% limit | Hooked a sig. better fit | Discr. Logn. a sig. better fit |
|---|---|---|---|---|---|---|
| Cancer Research | 9994 | 5.45 | 7.38 | 9.11 | 1000 | 0 |
| Computational Mechanics | 7776 | 4.29 | 6.22 | 7.77 | 1000 | 0 |
| Computer Science Applications | 8148 | 1.34 | 3.21 | 4.89 | 911 | 0 |
| Control and Optimization | 1043 | 0.3 | 2.22 | 3.81 | 627 | 0 |
| Critical Care and Intensive Care Medicine | 2625 | 0.95 | 2.73 | 4.13 | 823 | 0 |
| Cultural Studies | 4848 | 0.97 | 2.53 | 4.08 | 750 | 0 |
| Developmental Neuroscience | 1394 | 3.05 | NA* | 6.75 | 999 | 0 |
| Economics and Econometrics | 9974 | 1.51 | 3.49 | 5.41 | 949 | 0 |
| Energy Engineering and Power Technology | 7833 | 3 | 4.54 | 6.05 | 1000 | 0 |
| Filtration and Separation | 3282 | 0.86 | 2.71 | 4.28 | 802 | 0 |
| Food Science | 9992 | 7.72 | 9.61 | 11.31 | 1000 | 0 |
| Geochemistry and Petrology | 8292 | 8.07 | 9.95 | 11.67 | 1000 | 0 |
| Global and Planetary Change | 834 | -0.15 | 1.91 | 3.52 | 478 | 0 |
| Health  social science | 4352 | 1.78 | 3.73 | 5.4 | 962 | 0 |
| Health Information Management | 697 | -0.94 | 1 | 2.48 | 95 | 2 |
| Management Science & Operations Res. | 3993 | 2.32 | 4.07 | 5.81 | 986 | 0 |
| Marketing | 2260 | 0.47 | 2.41 | 4.15 | 687 | 0 |
| Metals and Alloys | 9964 | 1.86 | 3.47 | 4.9 | 967 | 0 |
| Neuropsychology & Physiological Psychology | 2927 | 2.67 | 4.46 | 6.01 | 999 | 0 |
| Nuclear and High Energy Physics | 9994 | 2.34 | 4.21 | 6.08 | 991 | 0 |
| Pharmaceutical Science | 9228 | 0.92 | 2.73 | 4.5 | 792 | 0 |
| Physical and Theoretical Chemistry | 9986 | 6.1 | 7.75 | 9.42 | 1000 | 0 |
| Virology | 6534 | 9.44 | 11.34 | 13.1 | 1000 | 0 |

*The hooked power law did not always fit the simulated samples from this subject (see section 4.2).

**Table 10**. Vuong test z value and 95% confidence intervals from 1000 repetitions of sets of 500 simulated citation counts generated from the hooked power law distribution. For each subject and each set of 1000 samples, the data was randomly drawn from a hooked power law distribution with α and B parameters that are optimal for the subject. The last two columns record the number of tests for which one distribution is a statistically significantly better fit than the other at the $p = 0.05$ level.

| Subject that parameters were taken from | Sample size (articles) | Lower 95% limit | Median Vuong | Upper 95% limit | Hooked a sig. better fit | Discr. Logn. a sig. better fit |
|---|---|---|---|---|---|---|
| Cancer Research | 500 | 2.42 | 3.91 | 5.02 | 992 | 0 |
| Computational Mechanics | 500 | -1.03 | 1.09 | 2.81 | 178 | 0 |
| Computer Science Applications | 500 | -0.24 | 1.52 | 3.04 | 293 | 0 |
| Control and Optimization | 500 | -1.97 | -0.02 | 1.96 | 25 | 25 |
| Critical Care and Intensive Care Medicine | 500 | -7.36 | -4.76 | -2.36 | 0 | 987 |
| Cultural Studies | 500 | -2.95 | -1.45 | 0.2 | 0 | 274 |
| Developmental Neuroscience | 500 | NA* | NA* | 3.32 | 281 | 0 |
| Economics and Econometrics | 500 | -0.79 | 0.95 | 2.38 | 98 | 1 |
| Energy Engineering and Power Technology | 500 | -7.55 | -5.34 | -3.02 | 0 | 997 |
| Filtration and Separation | 500 | -0.6 | 1.13 | 2.51 | 149 | 0 |
| Food Science | 500 | 2.44 | 3.91 | 5.18 | 992 | 0 |
| Geochemistry and Petrology | 500 | -1.13 | 0.83 | 2.8 | 137 | 1 |
| Global and Planetary Change | 500 | 0.37 | 2.2 | 3.53 | 622 | 0 |
| Health  social science | 500 | -0.42 | 1.57 | 3.35 | 317 | 1 |
| Health Information Management | 500 | 0.22 | 1.97 | 3.16 | 504 | 0 |
| Management Science & Operations Res. | 500 | -0.32 | 1.57 | 3.03 | 324 | 0 |
| Marketing | 500 | 0.14 | 1.84 | 3.18 | 446 | 0 |
| Metals and Alloys | 500 | -3.96 | -2.56 | -0.89 | 0 | 782 |
| Neuropsychology & Physiological Psychology | 500 | 2.1 | 3.64 | 4.97 | 983 | 0 |
| Nuclear and High Energy Physics | 500 | -0.16 | 1.42 | 2.86 | 255 | 0 |
| Pharmaceutical Science | 500 | -2.89 | -0.62 | 1.3 | 4 | 106 |
| Physical and Theoretical Chemistry | 500 | -0.66 | 1.35 | 3.13 | 275 | 0 |
| Virology | 500 | NA* | NA* | 3.75 | 354 | 0 |

*The hooked power law did not always fit the simulated samples from this subject (see section 4.2).

## 5.5 RQ5: Could all subjects have the same scale parameter when the discretised lognormal distribution is fitted?

Figure 4 shows that the scale parameters for the discretised lognormal distributions fitted to citation counts to journal articles from a single Scopus subcategory and year are substantially different from each other. The relatively narrow confidence intervals in the graph suggest that the differences in scale parameters are fundamental to the fields rather than due to uncertainty in the parameters when fitting the distribution. Hence the claim of a universal scale parameter (Radicchi, Fortunato, & Castellano, 2008) should be decisively rejected for Scopus subcategories. Nevertheless, it is possible that varying scale parameters could be due to mixing subfields with different location parameters within a single field (see online supplement S1) rather than to underlying differences in scale parameters for uniform fields. Uniform fields are probably rarely attainable in most citation analyses in practice due to interdisciplinary research and multiple specialisms within fields. The remaining theoretical possibility of a uniform scale parameter is therefore not testable and may not be relevant to citation distribution fitting.

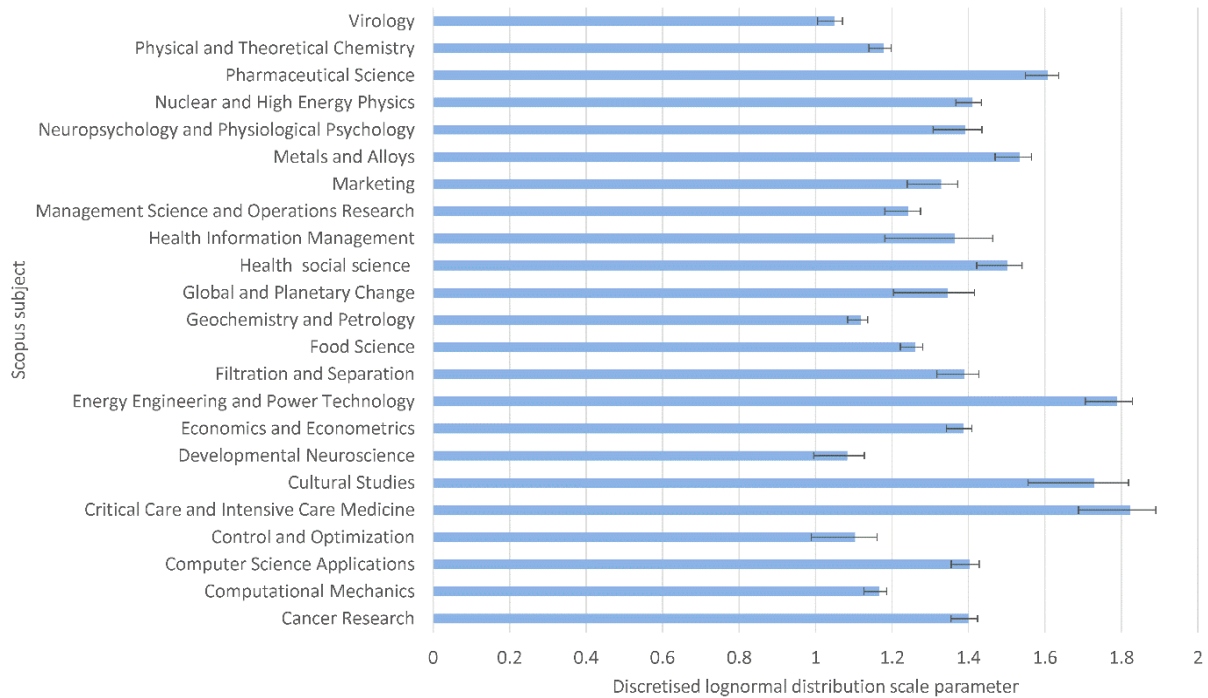

**Figure 4**. Scale parameters from discretised lognormal distributions fitted to citation counts to journal articles from a single Scopus subject in 2006. Error bars show 95% confidence intervals around the median from 1000 bootstrapped samples for each subject.

## 6. Discussion

The results are limited in that they do not cover all academic fields and years. In particular, data that is substantially newer may give different results (e.g., Šubelj & Fiala, in press). The Scopus field categorisation scheme is also a limitation because journal classification is subjective (see also: Börner, Klavans, Patek, et al., 2012). Much research is also interdisciplinary, which affects its citation counts (Larivière & Gingras, 2010), and articles may not fit into a specific subject and may also be published in journals that are classified outside of the subject area of the main article topic. Hence, the poor fit of a distribution may be due to the classification process rather than because the distribution does not fit the subject area and it is even possible that one of the two distributions fits better than the other because of the combination of subfields within a subcategory. The choice of journals to include in Scopus is also important because the list is not exhaustive and the selection criteria may be subjective and influenced by commercial considerations. Other factors may also influence the purity of a distribution in any particular set of articles, such as national differences in research performance, different research orientations (pure/applied), different levels of teamwork, and all factors that affect the citation process (Borgman & Furner, 2002).

The interpretation of the term plausible in terms of the Kolmogorov-Smirnov test is also a limitation because in practice, real datasets cannot be expected to ever exactly fit theoretical distributions due to real world limitations. In the current case, there may be individual recording and classification errors within Scopus or occasional anomalous bursts of publishing or citation practices (e.g., journal special issues, temporary issues with journals, or mid-year journal direction changes) and so it would not be reasonable to expect an exact fit. Moreover, as sample sizes increase, goodness of fit tests become more powerful and so minor deviations from an exact fit could lead to a distribution being rejected. Nevertheless, the systematic differences found (see Table 1 and the discussion below) give some confidence that the rejection of the plausibility of the hooked power law and discretised lognormal distributions is reasonable.

The results showed that for some disciplines the hooked power law was a statistically significantly better fit than the discretised lognormal distribution but for other fields the reverse was true. The *Articles from 2005 to 2014* data set was used to investigate whether this was a characteristic of the subject area or if the date was important. In all of the five subject areas one of the two distributions was a significantly better fit in most years, but not for all years (the full results are online: see Appendix). This suggests that the subject area is important for the shape of the citation distribution but is not the only factor.

One of the reasons that neither of the distributions tested is plausible for most of the subject subcategories is that Scopus subject categories may contain articles from multiple different subject subcategories, each following a different citation distribution. Mixing distributions reduces the extent to which a statistical model fits and affects the parameters of the fit. The root cause of the lack of a plausible fit by a distribution to the real data sets could therefore be that the Scopus subject subcategories are mixed rather than pure (graph in online supplement S1– see Appendix). Some degree of mixing within subjects is inevitable due to the presence of interdisciplinary research. Additional experimental simulation results confirm that mixed distributions fit less well than pure distributions (graphs in online supplement S2– see Appendix). They also show that it is possible with large sample sizes and substantial differences in means to detect with the Kolmogorov-Smirnov statistic that a dataset is combined rather than pure.

## 7. Conclusions

Taken together the results confirm that citation counts for a single subject and year do not follow a single universal discretised lognormal distribution, after compensating for different location parameters. The main problems with the universal lognormal distribution hypothesis (Radicchi, Fortunato, & Castellano, 2008) are that it does not have quite the right shape (Table 1), the distribution is statistically significantly different for most subjects (Table 2 – although this is not strong evidence because it is unrealistic to expect an exact match), the hooked power law fits statistically significantly better for several subjects (Tables 4,5) and, the scale parameters are statistically significantly different from each other for many pairs of subjects (Figure 4).

More generally, the results also give evidence that neither the hooked power law nor the discretised lognormal distributions are plausible for of articles from a single Scopus subject subcategory and year in the majority of cases because they both tend to have the wrong shapes (Table 1) and the distributions are statistically significantly different from the data for most subjects (Table 2 – although this is again not strong evidence). Whilst for some subject subcategories, one of the two distributions fits better than the other (Tables 4,5), neither is a universally better fit. The results also show that that in most cases the two distributions are not interchangeable in practice. An important implication is that if in future research a single distribution is used to model citation counts in general or in all fields then caution should be exercised when interpreting the results because it is likely that the distribution chosen is not a good fit for some subject areas. This most likely applies to the discretised lognormal distribution because the continuous lognormal distribution has analytic properties that make it more useful than the hooked power law for modelling citation data.

As mentioned in the Discussion above, the imperfect fits for both distributions in most subject subcategories could be due to Scopus subject categories being impure in the sense of containing articles from multiple sub-fields, each with its own citation distribution. In practice, detecting whether a real data set is combined is likely to be difficult because it may be constructed from more than two separate distributions with unknown sizes. Hence, the failure of the Kolmogorov-Smirnov tests (Tables 2,3) could either be due to the distributions not fitting or the subject subcategories being impure. It is therefore possible that one of the two distributions fits citation distributions well, in theory, even if they do not in practice. To test this hypothesis, it would be necessary to fit the distributions to sets of articles that are known to be pure in the sense of being narrowly focused on a specific topic and containing little interdisciplinary research.

In conclusion, although the results are conclusive in proving that typical Scopus subject subcategories cannot be plausibly claimed to universally fit either a discretised lognormal or hooked power law, this may be due to underlying mixing of distributions. It would be impossible to test for this on any given data set unless it is large and simplifying assumptions were made (e.g., that it was composed of exactly two similar discretised lognormal distributions with a common scale parameter). Nevertheless, both distributions fit sets of citation counts reasonably well and this conclusion does not rule out their use in practice. Returning to the rationale for this paper, because it is unlikely that one distribution can be identified as the correct distribution for citation counts, at least at a theoretical level, it will be difficult to use theoretical arguments based on citation count distributions in order to choose a particular citation indicator as having the most desirable properties. Instead, the relative merits of indicators could perhaps be assessed with both the hooked power law and the discretised lognormal distribution and if the properties are the same for both, then they are likely to also hold in practice.

## 9. Appendix

Online supplements S1 and S2, the code, processed data and figures for the main distributions are available online at: doi:10.6084/m9.figshare.3116656